\documentclass[twocolumn,prl,superscriptaddress,floats,nobibnotes,notitlepage,citeautoscript]{revtex4-1}

\usepackage[pdftex]{graphicx,hyperref}
\usepackage[utf8]{inputenc}
\usepackage{amsmath,bm}
\usepackage{bbold}
\usepackage{color,soul}
\usepackage{xstring}
\usepackage{wasysym}
\usepackage{xspace}
\usepackage{amssymb}
\usepackage{time}
\usepackage{bbold}
\usepackage{subfigure}
\usepackage{multirow}
\usepackage{xspace}
\usepackage{url}

\def\<{\langle}
\def\>{\rangle}
\DeclareMathOperator{\Tr}{Tr}

\renewcommand{\Im}[0]{\text{Im}\,}

\newcommand{\ve}[1]{\boldsymbol{#1}}

\makeatother

\definecolor{DarkRed}{RGB}{193,40,40}

\def\Tr{\mathop{\mathrm{Tr}}}

\makeindex

\begin{document}

\title{Doping-induced quantum spin Hall insulator to superconductor transition}

\author{Zhenjiu Wang }
\email{Zhenjiu.Wang@physik.uni-wuerzburg.de}
\affiliation{\mbox{Institut f\"ur Theoretische Physik und Astrophysik, Universit\"at W\"urzburg, 97074 W\"urzburg, Germany}}

\author{\firstname{Yuhai} \surname{Liu}}
\affiliation{\mbox{Beijing Computational Science Research Center, 10 East Xibeiwang Road, Beijing 100193, China}}
\affiliation{\mbox{Department of Physics, Beijing Normal University, Beijing 100875, China }}
\author{\firstname{Toshihiro} \surname{Sato}}
\affiliation{\mbox{Institut f\"ur Theoretische Physik und Astrophysik, Universit\"at W\"urzburg, 97074 W\"urzburg, Germany}}
\author{\firstname{Martin} \surname{Hohenadler}}
\affiliation{\mbox{Institut f\"ur Theoretische Physik und Astrophysik, Universit\"at W\"urzburg, 97074 W\"urzburg, Germany}}
\author{\firstname{Chong} \surname{Wang}}
\affiliation{\mbox{Perimeter Institute for Theoretical Physics,
Waterloo, Ontario, Canada N2L 2Y5}}
\author{\firstname{Wenan} \surname{Guo}}
\affiliation{\mbox{Department of Physics, Beijing Normal University, Beijing 100875, China }}
\affiliation{\mbox{Beijing Computational Science Research Center, 10 East Xibeiwang Road, Beijing 100193, China}}
\author{Fakher F. Assaad}
\email{assaad@physik.uni-wuerzburg.de}
\affiliation{\mbox{Institut f\"ur Theoretische Physik und Astrophysik, Universit\"at W\"urzburg, 97074 W\"urzburg, Germany}}
\affiliation{\mbox{W\"urzburg-Dresden Cluster of Excellence ct.qmat, Am Hubland, 97074 W\"urzburg, Germany}}

\begin{abstract}
A quantum spin Hall insulating state that arises from spontaneous symmetry breaking has remarkable properties:  Skyrmion textures of the SO(3)  order 
parameter carry charge 2$e$.   Doping this state of matter opens a new route to superconductivity via the condensation of  Skyrmions.   We define a model 
amenable to large scale negative sign free quantum Monte Carlo simulations  that allows us to study this transition.  Our results support a direct and continuous 
doping induced transition between the quantum spin Hall insulator and  s-wave superconductor. We can resolve dopings away from half-filling
down to $\delta = 0.0017$.   Such routes to superconductivity have been put forward in  the realm of twisted bilayer graphene.
\end{abstract}

\maketitle

{\it Introduction.}---  Doping a band insulator generically leads to a Fermi liquid state whose Fermi surface may become unstable to superconductivity.   In contrast,  insulating states  where correlation effects are dominant provide  different routes to superconductivity.  Low  lying Goldstone modes present in  the insulating state can provide a glue between doped charge carriers. Spin fluctuation theories of  high temperature superconductivity  follow this idea \cite{Coldea01,Scalapino12}.    The correlated insulator can also contain preformed pairs  that become charged upon doping.  The   resonating valence bond state  based theory of high temperature superconductivity follows this idea \cite{Anderson87,Lee06_rev}.  More recently, the idea of preformed pairs has been put forward in the realm of  graphene Moir\'{e}
superlattice systems~\cite{Zhangetal} such as twisted bilayer
graphene~\cite{Khalafetal}.   Here a  
correlation-induced topological insulator  contains 
Skyrmions  that carry charge 2e as low-lying excitations \cite{Grover08}. Upon doping superconductivity emerges due to the condensation of charged Skyrmions.

\begin{figure}
\centering
\includegraphics[width=0.45\textwidth]{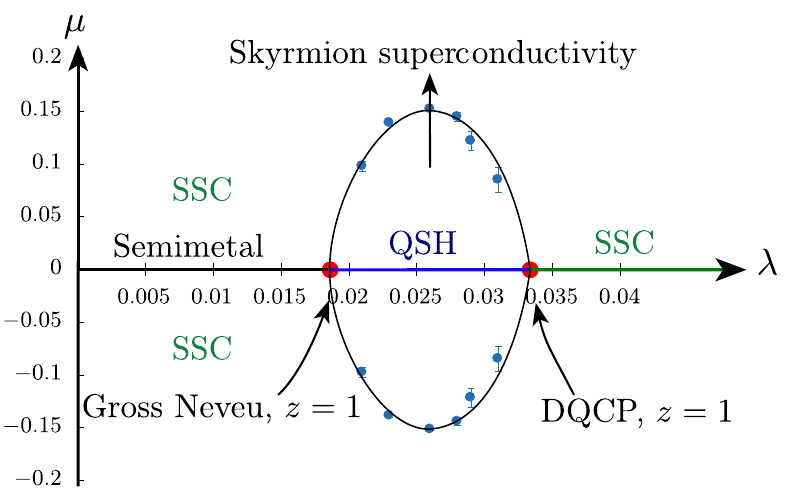}
\caption{\label{fig:Phase_diagram}
Phase diagram of the model of Eq.~\ref{Eq:Ham} in the interaction strength, $\lambda$, versus chemical potential  plane.  The data at half-filling is reproduced  from Ref.~\onlinecite{Liu_QSH}.  The critical chemical  potential $\mu_c$ at  which the transition from  the quantum spin Hall (QSH) to s-wave superconductor (SSC) occurs  is computed by measuring the pairing gap  at  half-filling (see Ref.~\onlinecite{SM} and Fig.~\ref{fig:Gap_Mom}(a)).
}
\end{figure}

The model we will considered in this Letter  differs significantly from the ones discussed in the realm of  graphene Moir\'{e}
superlattice systems but  captures  the essence of the aforementioned  topological route to superconductivity. 
In Ref.~\onlinecite{Liu_QSH}, we introduced a model of
Dirac fermions supplemented with a next-nearest-neighbor interaction term
($\sim\lambda$) and investigated its phase diagram at
half-filling (see Fig.~\ref{fig:Phase_diagram}). The interaction dynamically
generates a quantum spin Hall (QSH) insulating state that breaks SU(2) spin
rotational symmetry.  

Upon further increasing $\lambda$ at half-filling, the QSH state gives
way to an s-wave superconductor (SSC).  The QSH-SSC transition falls into the
class of deconfined quantum critical points (DQCPs)\cite{Grover08}. Our previous work
\cite{Liu_QSH} suggests that both phase transitions are described by conformal
field theories.  The semimetal to QSH transition is in the
Gross-Neveu-Yukawa universality class \cite{PhysRevD.10.3235} whereas the
DQCP is associated with a non-compact CP$^{1}$ \cite{senthil2004deconfined} field
theory describing the fractionalized SO(3) order parameter. 

The insulating QSH state has both preformed pairs, corresponding to Skyrmions
of the QSH order parameter, and Goldstone modes.  Understanding the fate of
this state as a function of doping is the aim of this Letter.  Our results
are consistent with a doping-induced weakly first-order or continuous QSH-SSC
transition 
driven by the condensation of Skyrmions.

\textit{Model and Method.}---We consider a model of Dirac fermions in $2+1$
dimensions on the honeycomb lattice with Hamiltonian
\begin{equation}\label{Eq:Ham}
\begin{aligned}
 \hat{H}  = - t  \sum_{ \langle \bm{i}, \bm {j} \rangle } (\hat{\ve{c}}^{\dagger}_{\bm{i} } \hat{\ve{c}}^{\phantom\dagger}_{\bm{j}}  + H.c.)
   -\lambda \sum_{\varhexagon}  \left( \sum_{\langle \langle \bm{i} \bm{j} \rangle \rangle  \in \varhexagon }   \hat{J}_{\bm{i},\bm{j}} \right)^2
\end{aligned}
\end{equation}
with
$ \hat{J}_{\bm{i},\bm{j}} = i \nu_{ \bm{i} \bm{j} }
\hat{\ve{c}}^{\dagger}_{\bm{i}} \bm{\sigma}
\hat{\ve{c}}^{\phantom\dagger}_{\bm{j}} + H.c.$ The spinor
$\hat{\boldsymbol{c}}^{\dag}_{\ve{i}} =
\big(\hat{c}^{\dag}_{\ve{i},\uparrow},\hat{c}^{\dag}_{\ve{i},\downarrow}
\big)$ where $\hat{c}^{\dag}_{\ve{i},\sigma} $ creates an electron at lattice
site $\ve{i}$ with $z$-component of spin $\sigma$. The first term
accounts for nearest-neighbor hopping. The second term is a plaquette
interaction involving next-nearest-neighbor pairs of sites and phase factors
$\nu_{\boldsymbol{ij}}=\pm1$ identical to the Kane-Mele model
\cite{KaneMele05b}, see also Ref.~\cite{Liu_QSH}.  Finally,
$\boldsymbol{\sigma}=(\sigma^x,\sigma^y,\sigma^z)$ correspond to the Pauli
spin matrices. We used the ALF (Algorithms for Lattice Fermions)
implementation~\cite{ALF17} of the well-established auxiliary-field quantum
Monte Carlo (QMC) method~\cite{Blankenbecler81,White89,Assaad08_rev}.  Because
$\lambda > 0$, we can use a real Hubbard-Stratonovich decomposition for the
perfect square term.  For each field configuration, time-reversal symmetry is
present, even at finite chemical potential, so that eigenvalues of the
fermion matrix occur in complex conjugate pairs.  Hence, we do not suffer from
the negative sign problem.  In contrast to Ref.~\cite{Liu_QSH}, we used a
projective version of the algorithm (PQMC)
\cite{Sugiyama86,Sorella89,Assaad08_rev}.  The PQMC is a canonical approach
in which the ground state is filtered out of a trial wave function that is
chosen to be a Slater determinant.  To avoid the negative sign problem, the
trial wave function has to be time-reversal symmetric, so that we can only
dope away from half-filling with Kramers pairs. For the considered trial wave
function (see Ref.~\cite{SM} for further details), we observed that a
projection parameter set by the linear length of the lattice is sufficient to
reach the ground state.

\begin{figure}
\centering
\includegraphics[width=0.45\textwidth]{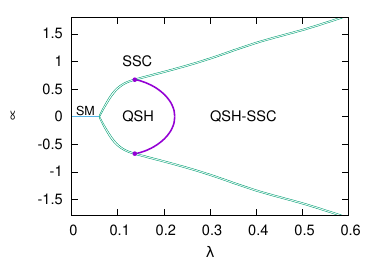}
\caption{\label{fig:MF_Phase_diagram}
Mean-field ground-state phase diagram. The blue and purple (green)  lines correspond to continuous (first-order) transitions.
}
\end{figure}

\textit{Mean-field approaches.}---Before discussing our QMC results,
it is instructive to carry out a mean-field approximation.  When expanding
the square in Eq.~(\ref{Eq:Ham}), diagonal terms,  $ \hat{J}_{\ve{i},\ve{j}}^2$,
contain, among other interactions, s-wave pair hopping terms that allow us to
introduce an SSC order parameter. The off-diagonal terms allow for QSH
ordering (see Ref.~\cite{SM} for a detailed calculation).  As seen in
Fig.~\ref{fig:MF_Phase_diagram}, doping the semimetal produces the SSC.
This reflects the pairing instability of Fermi surfaces to attractive
interactions within Bardeen-Cooper-Schrieffer (BCS) theory.
The protecting symmetries of the QSH state are related to time reversal and global charge conservation.
Hence, the coexistence region (QSH+SSC)  is  topologically  trivial. Furthermore, the transition at half-filling
 from the QSH to QSH+SSC is continuous and does not require the closing of the single-particle gap.
 Upon doping, the mean-field approximation generically supports two
 scenarios: (i) a continuous transition with dynamical exponent $z=2$ from
 the QSH to QSH+SSC, (ii) a first-order transition from the QSH to SSC
 \footnote{An intermediate metallic state would be unstable to pairing and we
   have excluded fine-tuning}.    Our mean-field approximation provides
 examples of both scenarios. As expected, it fails to capture the DQCP between
the QSH and SSC phases at half-filling \cite{Liu_QSH}.

{\it QMC results.}---We now turn to unbiased QMC results which, in
contrast to the mean-field approach, capture Goldstone modes as well as
topological Skyrmion excitations.  We consider $t=1$ and $\lambda = 0.026$, which
places us in the center of the QSH phase at half-filling.

\begin{figure}
\centering
\includegraphics[width=0.48\textwidth]{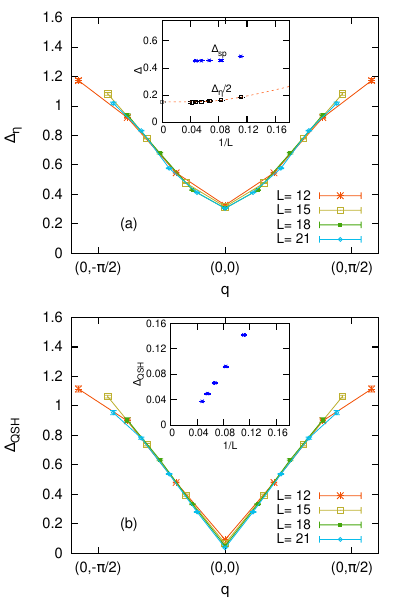}
\caption{\label{fig:Gap_Mom}
Momentum dependence of (a) the pairing gap  and (b) the QSH gap for the half-filled case,
 in the vicinity of $\Gamma$ point along the direction towards the $M$ point
 in the Brillouin zone of the honeycomb lattice.
 The inset of (a) shows the $1/L$ dependence of the single-particle gap
 $\Delta_{\text{sp}}$ and half of the s-wave pairing gap $\Delta_{\eta}/2$.
 The inset of (b)  shows the QSH gap, $\Delta_{\text{QSH}}$,  versus $ 1/L$.}
\end{figure}

At this filling, we show in Fig.~\ref{fig:Gap_Mom} the momentum dependence of
the spin-orbit coupling gap $\Delta_{\text{QSH}}(\ve q)$ and the SSC
gap $\Delta_{\eta}(\ve q)$. To obtain these data, we measured the imaginary-time
displaced correlation functions of the spin-orbit coupling operators
$\hat{\boldsymbol{O}}^{\text{QSH}}_{\boldsymbol{r},n}= \hat{J}_{\ve{r} +
  \ve{\delta}_n,\ve{r} + \ve{\eta}_n}$.  Here, $\ve{r}$ denotes a unit cell
and $n$ runs over the six next-nearest neighbor bonds of the corresponding
hexagon with legs $\ve{r} + \ve{\delta}_n$ and $\ve{r} + \ve{\eta}_n$.  We
also consider the s-wave pairing operators
$\hat{\eta}^{+}_{\ve{r},\ve{\tilde{\delta}}} = \hat{c}^{\dagger}_{\ve{r}
  +\ve{\tilde{\delta}},\uparrow} \hat{c}^{\dagger}_{\ve{r}
  +\ve{\tilde{\delta}},\downarrow}$, where $\tilde{\delta}$ runs over the two
orbitals in unit cell $\ve{r}$.  The gaps were obtained from
\begin{eqnarray}\label{Eq:gap}
  S^{\text{QSH}}(\ve{q},\tau)\nonumber
  &=&
  \sum_{n }  \langle  \hat{ \boldsymbol{O} }^{ \text{QSH} }_{\bm{q},   n } (
    \tau)   \hat{ \boldsymbol{O} }^{ \text{QSH} }_{- \bm{q}, n } (0) \rangle
    \propto
    e^{-\Delta_{\text{QSH}} (\ve{q}) \tau  }  \\\nonumber
  S^{\text{SSC}}(\ve{q},\tau)
  &=&
    \sum_{\ve{ \tilde{\delta} } }  \langle
                              \hat{\eta}^{+}_{\bm{q}, \ve{ \tilde{\delta} } }
                              (\tau)  \hat{\eta}^{-}_{ \bm{q}, \ve{
                              \tilde{\delta} } } (0)  +
                              \hat{\eta}^{-}_{\bm{q}, \ve{ \tilde{\delta} } }
                              (\tau) \hat{\eta}^{+}_{\bm{q}, \ve{
    \tilde{\delta} } } (0)   \rangle \\
 &\propto& e^{-\Delta_{\eta}(\ve{q}) \tau  },
\end{eqnarray}
in the limit of large imaginary time $\tau$ \cite{SM}.  As
expected for a Goldstone mode, $\Delta_{\text{QSH}}(\ve q)$ in
Fig.~\ref{fig:Gap_Mom}(b) exhibits a gapless, linear dispersion around the
ordering wave vector $\ve{q} = \Gamma$.  On the other hand,
$\Delta_{\eta}(\ve q)$ remains clearly nonzero with quadratic dispersion (see
Fig.~\ref{fig:Gap_Mom}(a)).  It is also important to note that an s-wave
pair has a smaller excitation energy than twice the single-particle gap,
as shown in the inset of Fig.~\ref{fig:Gap_Mom}(a).  Thus, pairing is present and
we can foresee that these preformed pairs will condense to form a
superconducting state upon doping.

\begin{figure}
\centering
\includegraphics[width=0.5\textwidth]{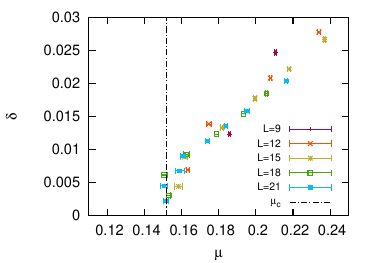}
\caption{\label{fig:Filling}
Doping factor $ \delta$ as a function of chemical potential
 $\mu  \equiv \frac{\Delta_{ \eta^{-} }}{2}$ for sizes $L=9$, 12, 15, 18, and $21$.
 The red dashed line is the critical chemical potential from the extrapolated
 pairing gap $\Delta_{\eta}/2 $ shown in Fig.~\ref{fig:Gap_Mom}.}
\end{figure}

A key quantity to understand the nature of the metal or superconductor to
insulator transition is the behavior of the chemical potential upon doping
away from half-filling \cite{Imada_rev,Imada_98_dop,Fisher89}.  For
first-order transitions,  $\mu$ shows a jump.  For continuous transitions, and with the assumption of a single length scale,
the singular part of the free energy scales as $f \propto |\mu -
\mu_c|^{\nu(d+z)}$ with $d$ the dimensionality and $\nu$
($z$) the correlation length (dynamical) exponent.  Since the doping defined as  $1-n$ is proportional to 
$\partial f/\partial \mu $ and the compressibility is
associated with twisting boundaries in the imaginary-time direction, one can show that for
transitions driven via the chemical potential  the hyper scaling
relation $\nu z = 1$ holds.  Thereby,
% \begin{equation}
 $ \delta  \propto  |\mu - \mu_c|^{ \nu d  }$,  
% \label{Eq:sl}
%\end{equation}
Doping a band insulator satisfies the hyper-scaling assumption. For a
quadratic band, $z=2$ so that $ \delta \propto |\mu - \mu_c|^{ d/2 } $.
This scaling behavior is satisfied upon doping a bosonic Mott insulator
\cite{Fisher89}.

\begin{figure}[h]
\centering
\includegraphics[width=0.48\textwidth]{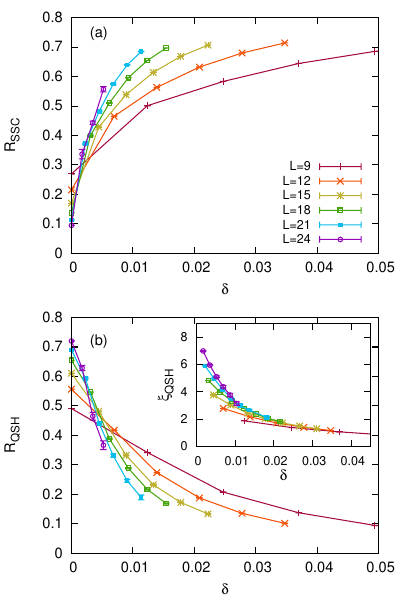}
\caption{\label{fig:Corr_Ratio}
Correlation ratios for (a) SSC and (b) QSH orders as a function of doping $\delta$.
The system sizes are $L=9$, 12, 15, 18, 21, and 24.
The inset of (b) shows the $\delta$-dependence of the finite-size correlation
length for the QSH order parameter.
}
\end{figure}

With the PQMC, we can compute the ground-state energy for a given, even
particle number $N_\text{p}$ and then derive the chemical potential. However, we found it
more efficient to extract $\mu$ from an estimate of
$\Delta_{ \eta^{-} }(N_\text{p})$ by analyzing the long imaginary time behavior of
the pair correlation function
$\sum_{\ve{\tilde{\delta}}} \langle \hat{\eta}^{+}_{\bm{q}, \ve{
    \tilde{\delta} } } (\tau) \hat{\eta}^{-}_{ \bm{q}, \ve{ \tilde{\delta} }
} (0) \rangle \sim e^{-\Delta_{\eta^{-}} \tau } $, where $\ve{q} = \Gamma$.
In particular,
\begin{equation}
  \begin{aligned}
   \mu  \equiv \frac{ E(N_\text{p}) - E(N_\text{p}-2) }{2}  = \frac{\Delta_{ \eta^{-} }(N_\text{p})}{2}\,.
  \end{aligned}
   \label{Eq:cp-pqmc}
\end{equation}
With the doping relative to half-filling defined as $ \delta \equiv 1 -
\frac{N_\text{p}-1}{2 L^2}$ \footnote{Here, $N_\text{p} -1$ is
 the thermal average of particle numbers, with the $N_\text{p}$ and the $N_\text{p} -2$ sector tuned to have the same ground-state energy
 at the chemical potential defined in Eq.~(\ref{Eq:cp-pqmc}).},
we  obtain  the data shown in Fig.~\ref{fig:Filling}. For
alternative ways of computing $\mu$ see the SM~\cite{SM}.

Figure~\ref{fig:Filling} plots $\delta$ as a function of $\mu$. The
vertical dash-dotted line corresponds to the critical chemical potential.
The data support a linear behavior for $\mu > 0.16$, but this form would
overshoot the critical chemical potential.  In a narrow window of dopings,
 $\delta < 0.01$, we observe a downturn in the functional form. Within
our precision, we can  offer two interpretations: a weakly first-order
transition or a continuous transition with dynamical exponent $z > 2$.  We note that
continuous metal-insulator transitions with $z>2$
have been put forward in the context of doped quantum antiferromagnets
\cite{Imada_rev,Assaad96}.

Another important question to answer is if the onset of superconductivity is
tied to the vanishing  of the QSH order
parameter.  To this end, we consider the renormalization-group invariant correlation ratios
($\alpha=\rm QSH, SSC$)
\begin{equation}
  R_{\alpha} \equiv  1 - \frac{S^\alpha(\bm{q}_{0}+\delta \bm{q})}{S^\alpha(\bm{q}_{0}) }
\end{equation}
based on the equal-time correlation functions of the spin current and s-wave paring
operators in momentum space, $S^{\alpha}(\mathbf q)$.  Here,
$\bm{q}_{0}=(0, 0)$ is the ordering wave vector and
$\bm{q}_{0}+\delta \bm{q}$ a neighboring wave vector.  By definition,
$R_{\alpha} \to 1$ ($\to 0$) in the ordered (disordered) state for
$L\to\infty$.  At a critical point, $R_{\alpha}$ is scale invariant and for
sufficiently large $L$, one should observe a crossing in $R_{\alpha}$ for
different system sizes.  Figures~\ref{fig:Corr_Ratio}(a) and (b) show results
for $R_{\rm SSC} $ and $R_{\rm QSH}$ as a function of $\delta$.  Due to the
observed binding of electrons in the insulating state, we expect
superconductivity for any $\delta>0$.  This is confirmed by
Fig.~\ref{fig:Corr_Ratio}(a).  The drift in the crossings due to corrections
to scaling is consistent with $\delta_c^{\rm SSC}\to 0$
in the thermodynamic limit.  The same quantity is plotted for the QSH
correlation ratio in Fig.~\ref{fig:Corr_Ratio}(b).  The data show that the
QSH order parameter vanishes very rapidly as a function of doping.
Again, the drift of the crossing point as a function of
  system size scales to smaller values of $\delta$.  Given the data, we can provide an upper bound
$\delta_c^{\rm QSH}< 0.0017$ which corresponds to our resolution \footnote{
  Since we are working in the canonical ensemble, the smallest doping is set
  by $2/(2L^2)$.}.  In our interpretation of Fig.~\ref{fig:Filling}, we could
not exclude the possibility of a weakly first-order transition.  On our finite systems, neither of
the correlation ratios show a discontinuity, consistent with a continuous transition.

As a crosscheck, we consider  the  second-moment, finite-size correlation length \cite{FPT_2015}, 
 $ {\xi}^2_{\alpha} \equiv  \frac{ \sum_{\boldsymbol{r}}  | \boldsymbol{r} |^2  S^{\alpha}(\boldsymbol{r})  }{ \sum_{ \boldsymbol{r}}  S^{\alpha}(\boldsymbol{r}) } $,  
obtained from the real-space, equal-time correlation functions
\footnote{The fact that there is no additional phase factor in the above summations comes from the known ordering wave vector $\boldsymbol{k}= \Gamma$.}.
The inset of Fig.~\ref{fig:Corr_Ratio}(b) reveals the absence of saturation of the QSH correlation length at any finite doping $\delta>0.0017$.
Saturation would be expected for a first-order transition.

\begin{figure}[h]
\centering
\includegraphics[width=0.48\textwidth]{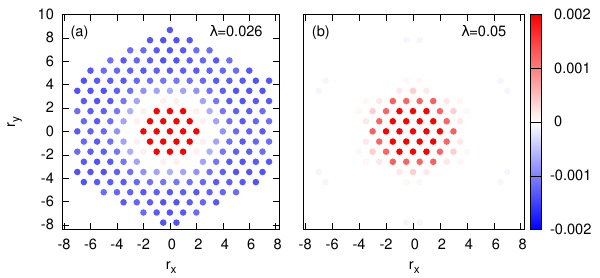}
\caption{\label{fig:Pinning} 
Real space equal time  correlation function of the  QSH order parameter around a pinned hole pair at the origin. 
To the Hamiltonian of Eq.~\ref{Eq:Ham}  we include the  pinning potential:
$ \hat{H}_{\text{pin}} \equiv C \sum_{\bm{r}} \sum_{ \tilde{\bm{\delta}} }
  e^{ -| (\bm{r} + \tilde{\bm{\delta}} ) | / \xi  } 
\hat{c}^\dagger_{\bm{r}+ \tilde{\bm{\delta}} } \hat{c}_{\bm{r}+ \tilde{\bm{\delta}} } $ 
with  $C=1$ and $\xi=1$. We consider 
$L=15$, at (a) $\lambda=0.026$ and (b) $\lambda=0.05$. 
}
\end{figure}

The notion of Skyrmion superconductivity hinges on a locking in of the charge   density  and  texture of the  SO(3) QSH order  parameter.    To image this, we dope  two holes away from half-filling  and localize  them  by modulating  the chemical potential.  The real space correlations  of the QSH order parameter are then expected to show a texture akin to a Skyrmion.  Precisely this is seen in  Fig.~\ref{fig:Pinning}(a)  at  $\lambda = 0.026$.   In contrast  \textit{ far } away from the QSH state at $\lambda = 0.05$  (see Fig.~\ref{fig:Pinning}(b))  a Skyrmion is not present around the localized pair.

{\it Discussion and summary.}---Our data suggest a doping-induced, continuous
and direct phase transition between the QSH state and the SSC.  Clearly, we
cannot exclude the possibility of a weakly first-order transition in which
the correlation length saturates beyond our maximum system size ($L=24$).
Our dynamically generated QSH state possesses Goldstone modes and charge-2$e$
Skyrmions of the QSH order parameter.  The Goldstone modes correspond to
long-wavelength fluctuations of the spin-orbit coupling and do not break
time-reversal symmetry. Hence, single-particle spin-flip scattering off
Goldstone modes---as present in doped quantum antiferromagnets---is not allowed.
Remarkably, one can also show that
$\left[ \hat{\ve{c}}_{\ve{k}=0} ,\hat{H}_\lambda \right] = 0$ (see Ref.~\cite{SM}), so
that at the $\Gamma$ point the single-particle spectral function \cite{SM} is
unaffected by the interaction $\hat{H}_\lambda$. This is in strong contrast
to quantum antiferromagnets, where Goldstone modes couple to single-particle
excitations to form a narrow band of spin polarons
\cite{Martinez91,Preuss95,Raczkowski2019}. These arguments suggest that
Goldstone modes do not provide the glue that leads to pairing.

We interpret our results in terms of preformed pairs, Skyrmions
carrying charge 2$e$, that condense upon doping.  In fact,  by pinning the charge we were able to image the Skyrmion. 
 Within this picture,
the correlation length that diverges at the transition corresponds to the
average distance between Skyrmions.

The finite-temperature phase diagram remains to be analyzed.  Such calculations could reveal pseudo-gap physics  related to preformed pairs at small
doping. At large dopings, a  crossover to  conventional superconductivity  is expected.

\begin{acknowledgments}
  FFA acknowledges many insightful discussions with M. Imada on the topic of
  metal-insulator transitions.  ZW would like to thank M. Ulybyshev and X. Wu
  for useful discussions on twisted bilayer graphene.  The authors gratefully
  acknowledge the Gauss Centre for Supercomputing e.V. (www.gauss-centre.eu)
  for funding this project by providing computing time on the GCS
  Supercomputer SUPERMUC-NG at Leibniz Supercomputing Centre (www.lrz.de).
  FFA thanks funding from the Deutsche Forschungsgemeinschaft under the grant
  number AS 120/15-1 as well as the W\"urzburg-Dresden Cluster of Excellence
  on Complexity and Topology in Quantum Matter ct.qmat (EXC 2147, project-id
  390858490).  ZW thanks financial support from the DFG funded SFB 1170 on
  Topological and Correlated Electronics at Surfaces and Interfaces.
  TS thanks funding from the Deutsche Forschungsgemeinschaft under the grant number SA 3986/1-1.
  Y.L. was supported by the China Postdoctoral Science Foundation
  under Grants No.2019M660432 as well as the National Natural Science Foundation of China
  under Grants No.11947232 and  No.U1930402.  Research at Perimeter Institute (C.W.) is supported by the Government of Canada through the Department of Innovation, Science and Economic Development Canada and by the Province of Ontario through the Ministry of Research, Innovation and Science. W.G. was supported by the National Natural
  Science Foundation of China under Grants No. 11775021 and No. 11734002.
\end{acknowledgments}

\bibliography{refs,fassaad}

\clearpage

\section{Supplemental material}

\maketitle

\subsection{Projective QMC approach}

We used the  projective QMC algorithm of the ALF-library \cite{ALF17}.
This canonical  algorithm  filters out the ground  state, $|\psi_0\rangle $,
from a trial wave function,  $|\psi_T  \rangle$,  that is required to be non-orthogonal to the ground state:
\begin{equation}
 \frac{  \langle \psi_0 |  \hat{O} | \psi_0  \rangle   }{\langle \psi_0 |   \psi_0  \rangle } = \lim_{\Theta \rightarrow \infty}
  \frac{ \langle \psi_T |  e^{- \Theta \hat{H}}  \hat{O}  e^{- \Theta \hat{H}}  | \psi_T \rangle }{
   \langle \psi_T |  e^{ - 2\Theta \hat{H} } | \psi_T \rangle   }.
\end{equation}

The trial wave function $ | \psi_T \rangle  $ is chosen to be  a Slater determinant  with
$N_\text{p}$ particles ( $ \hat{N} |\psi_T \rangle = N_\text{p} | \psi_T \rangle  $ ).    In particular,

\begin{equation}
|\psi_T \rangle =  |\psi_T^{\uparrow} \rangle   \otimes  |\psi_T^{\downarrow} \rangle
\end{equation}
with
\begin{equation}
  |\psi_T^{\sigma} \rangle \equiv   \prod_{n=1}^{N_\text{p}/2} \left(  \sum_{\bm{i}} \hat{c}^{\dagger}_{\bm{i},\sigma}  U_{ \bm{i}, n} \right)   | 0 \rangle.
\end{equation}
 $ U_{ \bm{i}, n } $ is the $n^{\text{th}}$  single-particle  eigenstate, ordered in ascending   energy eigenvalues,  of   the spinless fermion Hamiltonian
\begin{equation}
  \hat{H}  = - t  \sum_{ \langle \bm{i}, \bm {j} \rangle } (\hat{c}^{\dagger}_{\bm{i} } \hat{c}^{\phantom\dagger}_{\bm{j}}  + H.c.)
  +   \sum_{ \langle \bm{i}, \bm {j} \rangle }   \xi_{ \bm{i}, \bm{j} }
   (\hat{c}^{\dagger}_{\bm{i} } \hat{c}^{\phantom\dagger}_{\bm{j}}  + H.c.).
\end{equation}
The first term corresponds to the tight-binding Hamiltonian on the honeycomb lattice.
We require  the perturbing  hopping  matrix elements  $ | \xi_{ \bm{i}, \bm{j}} |  \ll  t $  and  $\Im {\xi_{\bm{i}, \bm{j} } } = 0 $.
The sign and modulus of  $\xi_{\bm{i}, \bm{j} }$ are chosen randomly so that all energy eigenvalues of the spinless Hamiltonian are non-degenerate.
Our trial wave function hence breaks  lattice and point group  symmetries.
Crucially, however,  time-reversal symmetry is present.   Since $\lambda > 0$    (see Eq.~(\ref{Eq:Ham})), we can decouple  the  interaction with a real Hubbard-Stratonovich  transformation  such that both the  imaginary time propagation and the  trial wave function  are invariant under  time reversal:
\begin{equation}
  T \alpha  \begin{pmatrix} \hat{c}_{\ve{i}, \uparrow} \\  \hat{c}_{\ve{i}, \downarrow}  \end{pmatrix}  T^{-1}
 =  \Bar{\alpha}    \begin{pmatrix} \hat{c}_{\ve{i}, \downarrow} \\ - \hat{c}_{\ve{i}, \uparrow}  \end{pmatrix}.
\end{equation}
Hence, the     eigenvalues of the fermion matrix  come in   complex conjugate pairs  and no  negative sign problem occurs.

A projection length $\Theta =L $ was found to be sufficient to converge to the finite-size ground state
for all of our system sizes.
We have used an imaginary time  step   $\Delta_{\tau}=0.2$
and  a  symmetric  Trotter  decomposition  to
guarantee the Hermiticity of the imaginary time propagator.  All calculations were carried out at  $\lambda= 0.026$   in units where $t=1$.

\subsection{Equal-time structure factor}

\begin{figure}[h]
\centering
\includegraphics[width=0.48\textwidth]{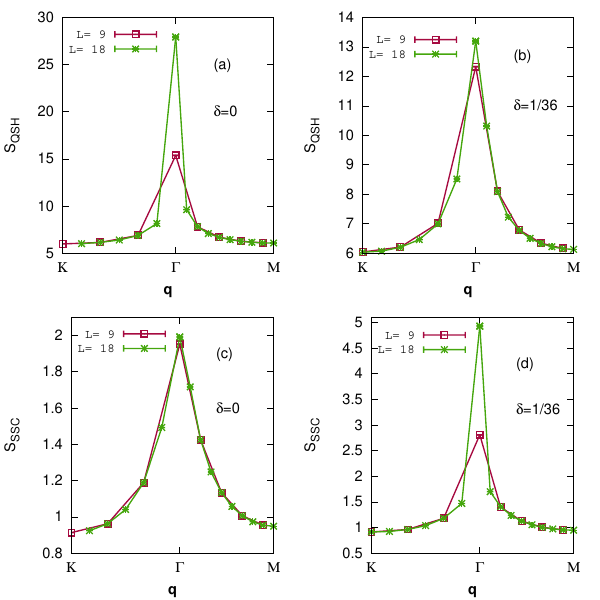}
\caption{\label{fig:S_K}
Momentum dependence of the equal-time structure factor for ((a),(b)) QSH and
((c),(d)) SSC operators for ((a),(c)) $\delta=0$ and
and ((b),(d))  $\delta=\frac{1}{36}$.
}
\end{figure}

In Fig.~\ref{fig:S_K}, we show the  momentum dependence of the  equal-time  QSH and SSC structure factors  at $\delta =0$ and  at
$\delta = 1/36$.   Upon doping, the QSH structure factor does not develop
incommensurate   features.   At $\delta = 1/36$,  the QSH data
(Fig.~\ref{fig:S_K}(b)) are consistent with the absence of long-range order,
whereas the SSC  structure factor (Fig.~\ref{fig:S_K}(d)) shows a marked
increase as a function of system size.

\begin{figure}[h]
\centering
\includegraphics[width=0.48\textwidth]{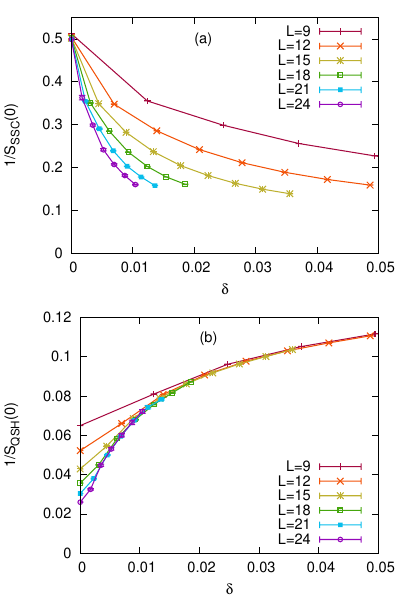}
\caption{\label{fig:S_delta}
$\delta$ dependence of $1/S(\bm{Q}=0)$ for SSC (a) and QSH (b), as a function of $\delta$, for $L=9$, 12, 15, 18, 21 and $24$}
\end{figure}

The onset of long-range order as well as a measure for the correlation length
can be obtained by considering $1/S(\bm{Q}=0)$ as function of $\delta$ (see
Fig.~\ref{fig:S_delta}).  The SSC ordering appears immediately at $\delta>0$,
characterized by the quick decay of $1/S_{\text{SSC}}$ as function of system
size. In particular, $1/S_{\text{SSC}}$ shows no saturation as a function of
system size.  On the other hand, $1/S_{\text{QSH}}$ shows a clear saturation
at \textit{large} doping.  For a given doping, the lattice
size at which this quantity converges is a measure of the correlation
length. Upon inspection, one will see that
larger lattice sizes are required to achieve convergence upon approaching half-filling.  In particular,
following the envelope of these curves again suggests that the correlation
length of the QSH fluctuations grows continuously and diverges as
$\delta \rightarrow 0$.  This is consistent with the data shown in the main
text.

\subsection{Consistency check of the pairing gap}

\begin{figure}
\centering
\includegraphics[width=0.48\textwidth]{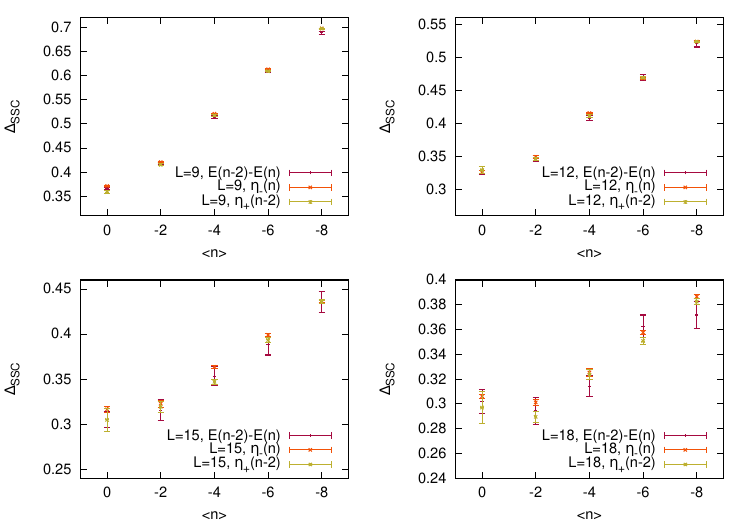}
\caption{\label{fig:Gap_Energy}
Three ways of pairing gap evaluation for (a) $L=9$, (b) $12$, (c) $15$, and (d) $18$, for doped particle number
$N_\text{p} - 2L^2 = 0, -2, -4, -6$ and $-8$.
}
\end{figure}

We check the consistency of our evaluation of  the ground-state energy difference between different even particle-number
sectors:
\begin{equation}
  E(N_\text{p}) - E(N_\text{p} -2) = \Delta_{ \eta^{-} }(N_\text{p})  = \Delta_{ \eta^{+} }(N_\text{p} -2)
\end{equation}
where $E(N_\text{p})$ is the ground state energy measured within the PQMC in the  $N_\text{p}$ particle number sector;
$ \Delta_{ \eta^{-} }(N_\text{p})$ is the s-wave pairing ($\eta^{-}$) gap extrapolated from
the time-displaced correlation function.

The imaginary-time domain $\beta$, in which we measure the time-displaced
correlation function, is set to $\beta=L$ for $L=9$, 12, 15, and $18$, and to $\beta=10$ for $L=21$.
To extrapolate the pairing gap, we use sequential fits
\begin{equation}
 \langle \eta^{+}( m \tau_0 + \tau ) \eta^{-} (m \tau_0)  \rangle  \propto  e^{ -\Delta_m \tau }   \  \  \   m = 0,1,2,3...
\end{equation}
where $\tau_0=1.0$ and $\tau \in [0, \tau_0)$.  The gap is extrapolated as
\begin{equation}
 \Delta_m - \Delta( m \rightarrow \infty )  \propto  e^{-a m}
\end{equation}
where $a$ is optimized for the best fit.

In  Fig.~\ref{fig:Gap_Energy}, we show that   the three different  ways of evaluating the  gap give  consistent results  for
$L=9$, 12, 15, and 18 for several particle-number sectors near half-filling.
In particular, one can compute the  ground-state energy and take the
difference or  measure time-displaced correlation functions of the pair
adding or removal operator.   Using the energy difference  generically
produces  bigger error bars. Here, we  carry out two independent simulations
and thereby  have to add the errors on two extensive  quantities (total
energies) to estimate the error on an intensive one, the total energy
difference.   Hence to keep the error bar on the total energy
difference, we have to scale the error on the energy per site as $1/L^2$.  Even
taking into account self-averaging on large system sizes, this proves to  be
numerically expensive.

\subsection{Gap extrapolation at half filling}

In this section  we discuss the calculations carried out to determine the  phase boundary of Fig.~\ref{fig:Phase_diagram}.  
As explained in the main text, the critical chemical potential is obtained from half  the pairing gap $\Delta_{\eta}$ at half filled case.   

\begin{figure}
\centering
\includegraphics[width=0.48\textwidth]{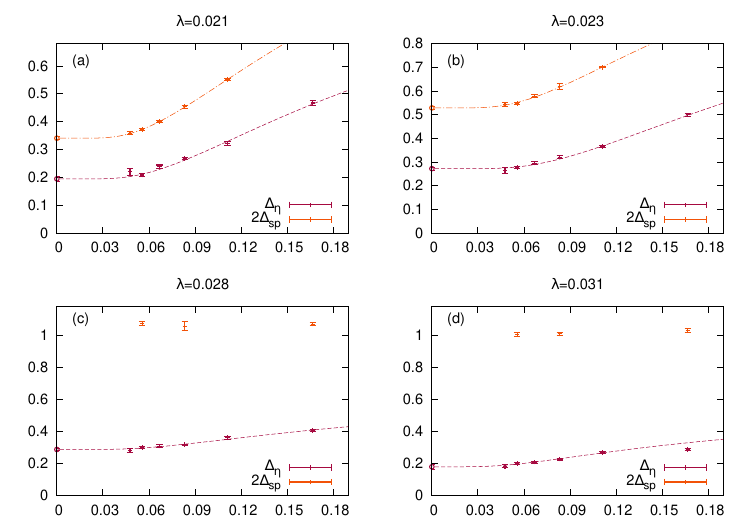}
\caption{\label{fig:Gap_Pair_Half_Filling} 
Pairing  and single particle gap extrapolation at half-filling,  for $\lambda=0.021$, $0.023$, $0.028$ and $0.031$.  
}
\end{figure}

Fig.~\ref{fig:Gap_Pair_Half_Filling} shows the extrapolation of  the  pairing gap $ \Delta_{\eta} $ at $\mu = 0$, for four different values of $\lambda$  
inside the  QSH phase.  An exponential finite size behavior is assumed for the extrapolation:  
\begin{equation}
    \Delta_{\eta} (L)  =   \Delta_{\eta} ( L \longrightarrow \infty )  + a e^{ -L / \xi } 
\end{equation}
For $\lambda=0.021, 0.023$ and $0.028$, an acceptable $\chi^2$ is obtained for a collective fit using sizes $L=6,9,...,21$.  
For $\lambda=0.031$,  an acceptable $\chi^2$ is obtained when the  $L=6$  data is omitted.

Additionally we show the single particle gap at $\mu=0$  as extracted from  the imaginary time displaced  Green function:    
\begin{equation}
\begin{aligned}
    \langle  c_{\bm{k} } (\tau)  c^{\dagger}_{ \bm{k}, \alpha }(0) \rangle  \propto  e^{ -\Delta_{ \text{sp} } \tau }. 
\end{aligned}
\end{equation}
Due to the shift of the minimal gap   in  momentum space  at   large $\lambda$,     
 we choose $ \bm{k}  \equiv ( \frac{4\pi}{3}, 0 ) $ for $ \lambda = 0.021, 0.023$ and $0.026$,  and $ \bm{k}  \equiv ( \pi, \frac{\pi}{ \sqrt{3}} ) $ for 
$ \lambda = 0.028$ and $0.031$.  

To extrapolate to the thermodynamic limit, we have again used an exponential fit,
\begin{equation}
    \Delta_{ \text{sp} } (L)  =   \Delta_{ \text{sp} } ( L \longrightarrow \infty )  + a e^{ -L / \xi },
\end{equation}
that is supported by the data. 

Before proceeding, we would like to comment on the exponential fit, since it is unexpected. In principle,  a doped hole (or pair)  will couple to the Goldstone modes  
such that  one expects the finite size effects of the gap to pick up the finite size  behavior of the Goldstone mode.  The latter follow a polynomial law  in $1/L$.  
This behavior is explicitly seen in the context of antiferromagnetic Mott insulators \cite{Assaad96a}.   If on the other hand the doped  charge carriers do not couple 
strongly to the low-lying  Goldstone modes,  then we expect dominant exponential finite size effects.  This point of view will be confirmed  in the  next section.

\subsection{Single-particle spectrum at finite doping}

\begin{figure}
\centering
\includegraphics[width=0.50\textwidth]{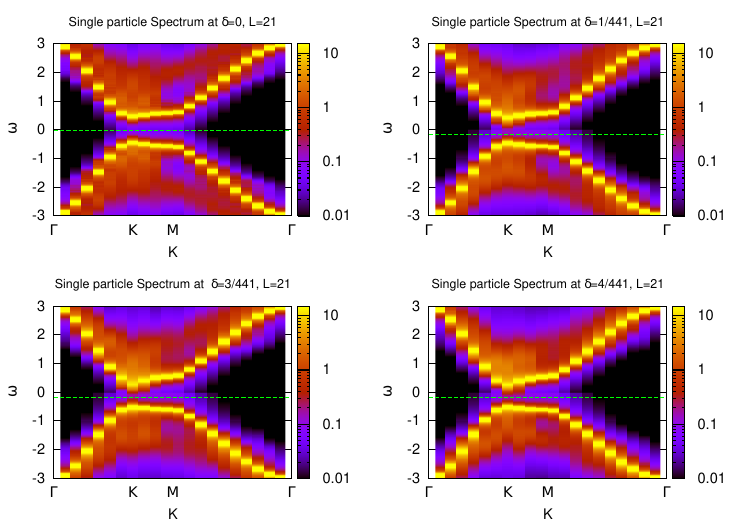}
\caption{\label{fig:Single_particle_spectrum}
Single-particle spectrum at dopings  (a) $\delta=0$, (b)
$\frac{1}{441}$, (c) $\frac{3}{441}$, and (d) $\frac{4}{441}$.
The green dotted line is the  chemical potential $ \mu $ evaluated from
Eq.~(\ref{Eq:cp-pqmc}) of the main text.
}
\end{figure}

We consider the single-particle spectral function at finite doping.    Away
from half-filling,  particle-hole symmetry is broken  and we have to
separately calculate the spectra for electron addition and removal,
\begin{equation}
\begin{aligned}\label{Eq:Spectrum}
A(\bm{k}, \omega) & = \frac{1}{Z}  \sum_n
  ( | \langle  n | c_{\bm{k}} | 0 \rangle |^2  \delta(E_n - E_0 - \omega)  )  \\
& +  \frac{1}{Z}  \sum_m ( | \langle  m | c^{\dagger}_{\bm{k}} | 0 \rangle |^2  \delta(E_m - E_0 + \omega)  )
\end{aligned}
\end{equation}
via the independent analytical continuations
\begin{equation}
\begin{aligned}
  &  \langle  c_{\bm{k} } (\tau)  c^{\dagger}_{ \bm{k}, \alpha }(0) \rangle  =  \int d\omega  e^{ -\tau \omega }  A_{+}(\omega )      \\
  &  \langle  c^{\dagger}_{\bm{k} } (\tau)  c_{ \bm{k}, \alpha }(0) \rangle  =  \int d\omega  e^{ -\tau \omega }  A_{-}(\omega )
\end{aligned}
\end{equation}
with $A(\omega)=A_{+}(\omega) + A_{-}(-\omega)$.  Here, $|0 \rangle$ in
Eq.~(\ref{Eq:Spectrum}) is the ground state at finite doping and $\langle n |$
is an eigenstate of the Hamiltonian with energy $E_n$ and an additional particle
(hole) relative to the ground state.  In
Fig.~\ref{fig:Single_particle_spectrum}, we plot the spectral functions for $L=21$ and
$\delta=0, \frac{1}{441}, \frac{3}{441}$ and $\frac{4}{441}$
($2L^2-N_\text{p} =0,2,6$ and $8$).  The dominant feature follows the
mean-field BCS form $E(\ve{k}) = \pm \sqrt{ (\epsilon(\ve{k}) -\mu )^2 + |\Delta|^2} $, where
$\pm \epsilon(k)$ denotes the Dirac dispersion of the honeycomb lattice.
This result shows that the Goldstone modes do not strongly couple to single-particle excitations.

In fact, the Green's function at the $\Gamma$ point has a special property,
due to a commutation rule between the fermion operator and the interaction
term of Hamiltonian (we use the notation $ \ve{c}^{\dagger}_{\ve{i}} =  ( \hat{c}^{\dagger}_{\bm{i}, \uparrow },\hat{c}^{\dagger}_{\bm{i}, \downarrow } )$)
\begin{equation}\label{Eq:Conserve_Int}
\begin{aligned}
\left[ \sum_{ \bm{i} }  \hat{c}_{\bm{i}, \alpha }  ,  \sum_{\varhexagon}  \left( \sum_{\langle \langle \bm{i} \bm{j} \rangle \rangle
\in  \varhexagon }  i \nu_{ \bm{i} \bm{j} }
   \hat{\ve{c}}^{\dagger}_{\bm{i}} \bm{\sigma} \hat{\ve{c}}^{\phantom\dagger}_{\bm{j}}  + H.c. \right)^2    \right]      =  0.
\end{aligned}
\end{equation}
The above relation follows directly from
\begin{equation}\label{Eq:Conserve_Hexagon}
\begin{aligned}
\left[ \sum_{ \bm{i} }  \hat{c}_{\bm{i}, \alpha }  ,    \sum_{\langle \langle \bm{i} \bm{j} \rangle \rangle  \in  \varhexagon  }  i \nu_{ \bm{i} \bm{j} }
   \hat{\ve{c}}^{\dagger}_{\bm{i}} \bm{\sigma} \hat{\ve{c}}^{\phantom\dagger}_{\bm{j}}  + H.c.      \right]      =  0
\end{aligned}
\end{equation}
which holds for the summation of spin-orbit  operators inside each hexagon
and for any vector $\ve{\sigma}$ in the space of Pauli matrices $\sigma_x$, $\sigma_y$, $\sigma_z$.
Hence, the Green's function  $\langle  \hat{\ve{c}}^{\dagger}_{\bm{k}} (\tau)  \hat{\ve{c}}^{}_{ \bm{k} }(0) \rangle$  at the  $\bm{k} = \Gamma$ point is identical to that
of the non-interacting Hamiltonian.

\subsection{Finite-temperature calculation}

\begin{figure}
\centering
\includegraphics[width=0.50\textwidth]{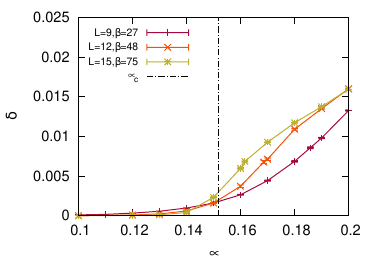}
\caption{\label{fig:Dop_FT}
Doping factor $ \delta $ as a function of chemical potential $\mu$ from FTQMC
simulations, with sizes $L=9$, $12$ and $15$, and $\beta=\frac{1}{3}L^2$.
}
\end{figure}

In contrast to the  projective  approach, the finite-temperature auxiliary
field QMC (FTQMC) algorithm is formulated in the grand-canonical ensemble.
Adding a chemical potential  term, $-\mu   \sum_{ \bm{i} } \hat{\ve{c}}^{\dagger}_{\bm{i} } \hat{\ve{c}}^{\phantom\dagger}_{\bm{i}} $,  to  the Hamiltonian, see  Eq.~(\ref{Eq:Ham}) of the
main text,  allows us to compute the doping  away from half-filling ($\mu = 0$):
 \begin{equation}
  \delta \equiv  \frac{ \langle \sum_{\bm{i}}  \hat{n}_{\bm{i}}  \rangle }{2 L^2} - 1.
 \end{equation}
Figure~\ref{fig:Dop_FT} shows the corresponding result for the case where the
inverse temperature $\beta$ and the system size $L$  scale  as $\beta=L^{2} /3$.
This implicitly  makes the assumption that $z=2$. Overall, our limited data
are consistent with the more efficient PQMC calculation.  At large values of
$\delta$, results for $L=12$ and $L=15$ are consistent with a linear dependence of
$\delta$ on $\mu$ that overshoots the critical chemical potential and suggest $z > 2$.

From the numerical point of view, the   FTQMC    is not as  efficient as the
PQMC.    The numerical cost to reach the low-temperature limit  scales as
$V^{3} \beta^{z} $.    We have also noticed  long warm-up  and autocorrelation
times  to equilibrate the particle number in the vicinity of $\mu = \mu_c $.

\subsection{Mean-field calculation}

In this section, we summarize the details of our mean field calculation.
Expanding interacting part of Eq.~(\ref{Eq:Ham}) of the main text as
\begin{equation}
\begin{aligned}\label{Eq:MF_decouple}
 H_V = &  -\lambda \sum_{\varhexagon}  \left(\sum_{\langle \langle \bm{i} \bm{j} \rangle \rangle }  i \nu_{ \bm{i} \bm{j} }
   \hat{c}^{\dagger}_{\bm{i}} \bm{\sigma} \hat{c}_{\bm{j}}  + H.c. \right)^2    \\
   = &  -\lambda  \sum_{\varhexagon}  \sum_{ \langle \langle \bm{i} \bm{j} \rangle \rangle }
   \sum_{ \langle \langle \bm{i'} \bm{j'} \rangle \rangle \neq  \langle  \langle \bm{i} \bm{j} \rangle \rangle  }
  \bm{ \hat{J}_{ \langle \langle i, j \rangle \rangle }  \cdot  \hat{J}_{ \langle \langle i', j' \rangle \rangle }  }    \\
   & - \lambda \sum_{\varhexagon}  \sum_{ \langle \langle \bm{i} \bm{j} \rangle \rangle }
   [ + 6 \hat{\eta}^{\dagger}_{\bm{i}}  \hat{\eta}_{\bm{j}} + h.c  -4 \bm{ \hat{S}_i } \cdot \bm{ \hat{S}_j }   \\
   &  -5 \hat{n}_{\bm{i}} \hat{n}_{\bm{j}}  +  5 ( \hat{n}_{\bm{i}} + \hat{n}_{\bm{j}} )    ]
\end{aligned}
\end{equation}
\begin{figure}
\centering
\includegraphics[width=0.42\textwidth]{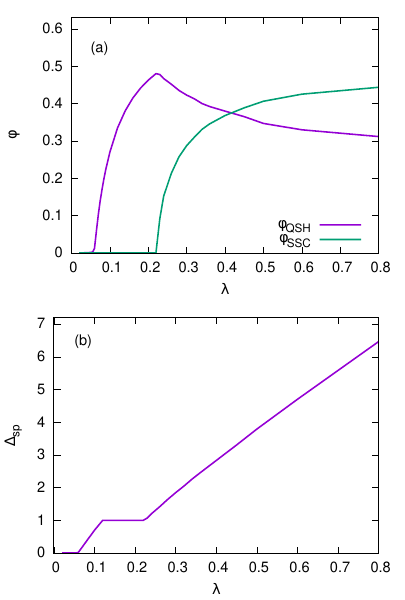}
\caption{\label{fig:MF_Half}
Mean-field solution as a function of $\lambda$ at half-filling. (a) QSH and SSC order parameters. (b) Fermionic single-particle gap.
}
\end{figure}
where
\begin{equation}
\begin{aligned}
 &  \bm{ \hat{J} }_{ \langle \langle i, j \rangle \rangle }  \equiv    i \nu_{ \bm{i} \bm{j} }
  \hat{c}^{\dagger}_{\bm{i}} \bm{\sigma} \hat{c}_{\bm{j}}  + H.c.,         \\
 &  \hat{\eta}_{\bm{i}}  \equiv  \hat{c}_{ \bm{i} \downarrow}  \hat{c}_{ \bm{i} \uparrow},     \   \   \    \
    \hat{\eta}^{\dagger}_{\bm{i}}  \equiv  \hat{c}^{\dagger}_{ \bm{i} \uparrow}  \hat{c}^{\dagger}_{ \bm{i} \downarrow},    \\
 &  \bm{ \hat{S}_i }  \equiv  \frac{1}{2}\hat{c}^{\dagger}_{\bm{i}}  \bm{\sigma} \hat{c}_{\bm{i}}.
\end{aligned}
\end{equation}
The self-consistent calculation is  based on selecting a polarization direction
for the three  (two) components of the QSH (SSC) order parameter.
The calculation is done by numerically minimizing the free energy in the space of the two order parameters
\begin{equation}
\begin{aligned}\label{Eq:MF_F}
  f(\beta)_{\phi}  =  \frac{-1}{\beta V} \ln{ \Tr{ e^{-\beta (H_T + H_V) - 15 \beta  V \lambda    {\phi_{\text{QSH}}}^2   -  36 \beta V \lambda    {\phi_{\text{SSC}}}^2    }  } }
\end{aligned}
\end{equation}
where
\begin{equation}
\begin{aligned}
 H_T = & - t  \sum_{ \langle \bm {i}, \bm {j} \rangle } (\hat{\ve{c}}^{\dagger}_{\bm{i} } \hat{\ve{c}}^{\phantom\dagger}_{\bm{j}}  + H.c.  )
   + \mu \sum_{\bm{i}} \hat{\ve{c}}^{\dagger}_{\bm i} \hat{\ve{c}}^{\phantom\dagger}_{\bm i}     \\
 H_V
   = &  - 5  \lambda  \sum_{\varhexagon}  \sum_{ \langle \langle \bm{i} \bm{j} \rangle \rangle }
     \phi_{\text{QSH}}
    \cdot  \hat{J^z}_{ \langle \langle i, j \rangle \rangle }       \\
   & -  36 \lambda   \sum_{ \bm{i} }
      \phi_{\text{SSC}}  \hat{\eta}^{x}_{\bm{i}}.
\end{aligned}
\end{equation}

\begin{figure}
\centering
\includegraphics[width=0.42\textwidth]{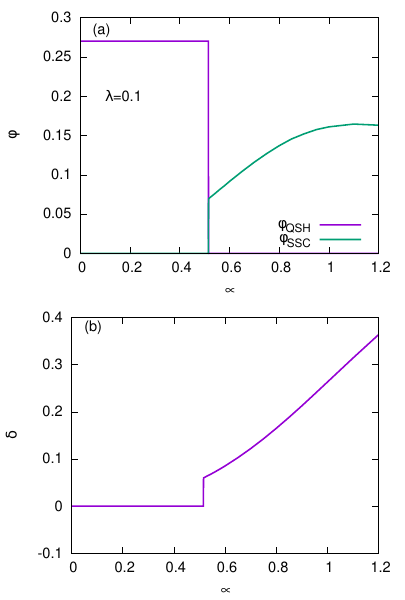}
\caption{\label{fig:MF_l0.1}
Mean-field solution as a function of chemical potential $\mu$ at $\lambda=0.1$. (a) QSH and SSC order parameters. (b) Doping factor $\delta$.
}
\end{figure}

\begin{figure}
\centering
\includegraphics[width=0.42\textwidth]{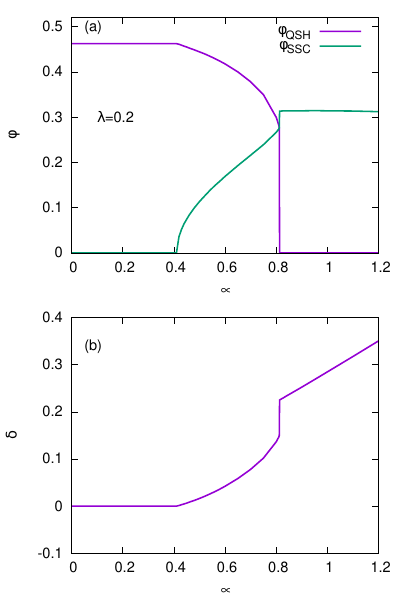}
\caption{\label{fig:MF_l0.2}
Mean-field solution as a function of chemical potential $\mu$ at $\lambda=0.2$. (a) QSH and SSC order parameters. (b) Doping factor $\delta$.
}
\end{figure}

We  consider a   paramagnetic saddle point  with   $ \langle \hat{\ve{S}}_{\ve{i}}  \rangle = 0 $  and $\langle \hat{n}_{\bm{i}} \rangle = 1 - \delta $.
Thus, for any local minimum of Eq.~(\ref{Eq:MF_F}) with $ \frac{\partial f}{
  \partial \phi_{\text{QSH}}} = 0 $ and  $ \frac{\partial f}{ \partial
  \phi_{\text{SSC}}} = 0 $,
\begin{equation}
\begin{aligned}
 & \phi_{\text{QSH}}  =  \frac{1}{6} \sum_{ \langle \langle \bm{i} \bm{j} \rangle \rangle }  \langle   \hat{J^z}_{ \langle \langle i, j \rangle \rangle }   \rangle  \\
 & \phi_{\text{SSC}}  =  \frac{1}{2} \langle  \hat{\eta}^{x}_{\bm{i},A} + \hat{\eta}^x_{\bm{i},B}    \rangle
\end{aligned}
\end{equation}
which holds locally due to  translational  symmetry.  We numerically integrated over the Brillouin zone of an $L=120$ lattice and took the zero-temperature limit $\beta \rightarrow \infty$.

The two order parameters as a function of $\lambda$ in the  half-filled case
are shown in Fig.~\ref{fig:MF_Half}(a).  We  observe a semimetal
($\phi_{\text{QSH}} = \phi_{\text{SSC}}=0$ ), a pure QSH  state
($\phi_{\text{QSH}} \neq 0, \phi_{\text{SSC}}=0$) as well as a coexistence
(QSH+SSC) state ($\phi_{\text{QSH}} \neq 0, \phi_{\text{SSC}} \neq 0$).
Since charge conservation is a  protecting symmetry of  the QSH  insulator,
the transition  between the QSH and QSH+SSC  states can be continuous without
a closing of the single-particle gap (see Fig.~\ref{fig:MF_Half}(b)).

On the other hand, upon doping the pure QSH state, the phase diagram  exhibits
two distinct mean-field scenarios.  Two representative examples at
$\lambda=0.1$ and $\lambda=0.2$ are shown in Figs.~\ref{fig:MF_l0.1} and
\ref{fig:MF_l0.2}, respectively.  In the case of $\lambda=0.1$
(Fig.~\ref{fig:MF_l0.1}), which is close to the Gross-Neveu transition, a
clear first-order transition between the QSH and SSC phases is observed.
Doping at $\lambda=0.2$ (Fig.~\ref{fig:MF_l0.2}) leads to two phase
transitions: (i) a $z=2$ transition from the pure QSH state to the coexistence
state at $\mu \approx 0.4$, characterized by a linear growth of $\delta$ and
(ii) a first-order phase transition to an SSC state at $\mu \approx 0.8$.  Such
first-order transitions are characterized by a level crossing corresponding
to two local minima in the free-energy density in Eq.~(\ref{Eq:MF_F}).

\end{document}